\documentstyle[preprint,aps]{revtex}
\draft
\begin{document}
\title{Revisit of Coulomb effects on $\pi^{-}/\pi^{+}$ ratio
in heavy ion collisions}
\bigskip
\author{\bf Bao-An Li}
\address{Cyclotron Institute and Department of Physics\\
Texas A\&M University, College Station, TX 77843, USA}
\maketitle

\begin{quote}
We show that the large $\pi^{-}/\pi^{+}$ ratio
at low pion energies recently observed in the reaction
of Au+Au at $E_{beam}/A=1.0$ GeV at SIS/GSI is due to
the strong Coulomb field of the reaction system.
\end{quote}

\newpage
It was found recently by the KaoS and FOPI collaborations at SIS/GSI
that the $\pi^{-}/\pi^{+}$ ratio increases as the pion kinetic energy
or transverse mass decreases \cite{baltes,muntzth,senger94,pelte94}.
The ratio is found to depend strongly on the mass of the reaction system.
For example, in the reaction of Au+Au at $E_{beam}/A=1.0$ GeV the
ratio increases continuously from about 1 at $E^{\pi}_{kin}\approx 400$ MeV
to about 2 at $E^{\pi}_{kin}\approx 100$ MeV. While in the reaction
of Ni+Ni at $E_{beam}/A=1.8$ GeV the ratio goes upto only
about 1.15\cite{oeschler}. More recently, the E866 collaboration
also studied the $\pi^{-}/\pi^{+}$ ratio as a function of the
pion transverse energy $E_t$ for Au+Au, Si+Au,
Si+Cu and Si+Al reactions at AGS/Brookhaven\cite{gonin}. For
the first two reactions the $\pi^{-}/\pi^{+}$ ratio rises to about 2
at low $E_{t}$, while for the last two systems the ratio is only about
1 in the whole energy range.

The phenomenon has renewed some interests on the study of $\pi^{-}/\pi^{+}$
ratio in relativistic heavy ion collisions. The main question of interest
is whether it is a manifestation of the well known Coulomb effect
or an indication of some new physics. Moreover, it seems also necessary to well
understand the isospin and Coulomb effects in order to study
the more interesting phenomena found in the pion spectra of
these experiments, such as, the enhancement of low/high energy pions.
For this purpose we perform a quantitative study on the $\pi^{-}/\pi^{+}$
ratio in heavy ion collisions at SIS/GSI energies using a hadronic
transport model \cite{li91a,li91b}. We show that the
large $\pi^{-}/\pi^{+}$ ratio found at SIS/GSI is indeed a manifestation
of the Coulomb effect. Although we consider here reactions at
SIS/GSI energies only, results of this study may also shed some light
on understanding the similar effect found at AGS energies.

For completeness we first mention some of the early studies
on the $\pi^{-}/\pi^{+}$ ratio in nuclear reactions.
A large $\pi^{-}/\pi^{+}$ ratio at very low laboratory energies
was observed a long time ago in experiments using protons,
alphas and cosmic rays, and was explained as due to the Coulomb
interactions. These studies were summarized and discussed in great detail
in ref. \cite{marshak}. Later on, much interest was generated by the
discovery of structures in charged pion spectra at the Bevalac.
A very distinctive feature observed in collisions of nearly equal mass
projectile and target nuclei is an enhancement of the $\pi^{+}$
yield in the mid-rapidity region, which appears as a peak
in the $\pi^{+}$ spectra at $\theta_{cm}=90^{0}$. Such a peak was
first observed at transverse momentum $p_{t}\approx 0.4 m_{\pi}c$
in the reaction of Ar+Ca at $E_{beam}/A=1.05$ GeV by Wolf et al.\cite{wolf79},
and at a slightly higher $p_{t}$ by
Chiba et al.\cite{chiba79} in the reaction of Ne+NaF at $E_{beam}/A=0.8$ GeV.
Additional structure in the low energy region at $\theta=0^{0}$ in the
projectile frame was found first by Benenson et al.\ \cite{benenson}
in the reaction of Ne+NaF at $E_{beam}/A\leq 400$ MeV.
Corresponding to the strong $\pi^{-}$ peak at $p_{t}=0$ a dip was seen
in the $\pi^{+}$ spectrum in the same experiments. Some systematic studies
on the $\pi^{-}/\pi^{+}$ ratio were also performed at the
Bevalac\cite{wolf82,radi82,schn82,sull82}

On the theoretical side, all studies seem to indicate the
Coulomb nature of the phenomenon. By comparing the $\pi^{+}, \pi^{-}$
wave functions in the electromagnetic field generated by the moving projectile
fragment, Bertsch was able to explain the peak in the $\pi^{-}/\pi^{+}$
ratio found by Benenson et al.\ and its beam energy
dependence\cite{benenson,bertsch80}. Libbrecht and Koonin showed
that the $\pi^{+}$ peak at $\theta_{cm}=90^{0}$
and $p_{t}=0.4 m_{\pi}c$ could be a result of the Coulomb focusing due to
the relatively moving charged fragments\cite{koonin79}. In their analysis
the charge distribution was approximated by spreading part of the
protons along a line between the nuclear centers with the remainder
continuing with the unaltered velocities. The classical equations
of motion for pions were then sloved in the Coulomb
field. Gyulassy and Kauffmann derived various approximate expressions for
$\pi^{-}/\pi^{+}$ ratio in the Coulomb field of projectile and target remanants
as well as the thermally expanding fireball \cite{gyulassy81}.
Later, the $\pi^{-}/\pi^{+}$ ratio was also studied in ref.\
\cite{bonasera} using a statistical model for pion production at
subthreshold energies\cite{nore88}.

In the present study we use the hadronic transport model \cite{li91a,li91b}
for relativistic heavy ion collisions. The model uses explicit isospin
degrees of freedom for all hadrons involved in the reaction. In particular,
isospin-dependent hadron-hadron collision cross sections and branching
ratios for the decay of baryon resonances are fully incorporated in the model.
The time dependent, selfconsist mean field for baryons including the
Coulomb field and an isosymmetry term is a basic ingredient of the model.
It has been rather successful in studying many aspects of
relativistic heavy ion collisions. We refer the reader to our
previous publications for more details of the model\cite{li94}.
In view of the fact that in most of the previous theoretical
studies on the subject one had to assume an idealized, simple and
often static charge distribution during the reaction, the present
study has the advantange of incorporating dynamically effects of the varying
charge distribution.

Fig.\ 1 shows the calculated $\pi^{-}/\pi^{+}$ ratio as a function of
pion kinetic energy with and without Coulomb interactions for the
reaction of Au+Au at $E_{beam}/A=1.0$ GeV.
In the case without Coulomb interaction the ratio is a constant
of about 1.9 for $E_{kin}\leq 500$ MeV, beyond 500 MeV the ratio has large
statistical error bars.
The constant $\pi^{-}/\pi^{+}$ ratio is very close to the estimate using
the isospin argument\cite{stock86}
\begin{equation}
R_{I}\equiv\frac{I^{-}}{I^{+}}=\frac{5N^{2}+NZ}{5Z^{2}+NZ},
\end{equation}
which gives 1.95 for the Au+Au reaction.
While in the case with Coulomb interactions the ratio is larger than the
isospin average for $E_{kin}\leq 150$ MeV and decreases as the pion energy
increases. The cross over of the two calculations at $E_{kin}\approx
150 $ MeV is a direct result of the competition between the Coulomb
momentum impulse and the phase space distortion.
This can be understood qualitatively using classical considerations
similar to that of refs.\ \cite{koonin79,gyulassy81}.
In terms of the inclusive cross section
$\sigma^{\pm}_{0}\equiv d^{3}\sigma^{\pm}_{0}/dp^{3}$
in the absence of the Coulomb field and the change of the phase
space density $\partial^{3}p_{0}/\partial^{3}p$,
the cross section in the presence of the Coulomb field
$\sigma^{\pm}\equiv d^{3}\sigma^{\pm}/dp^{3}$ can be written
as
\begin{equation}\label{csigma}
\sigma^{\pm}(\vec{p})=\sigma^{\pm}_{0}(\vec{p}_{0}(\vec{p}))
|\frac{\partial^{3}p_{0}}{\partial^{3}p}|\approx I^{\pm}\cdot\sigma_{0}
(\vec{p}_{0}(\vec{p}))
|\frac{\partial^{3}p_{0}}{\partial^{3}p}|,
\end{equation}
where $\vec{p}$ is the momentum of pions after moving through the Coulomb
filed starting with an initial momentum $\vec{p}_{0}$. In the last
approximation we factorized out the isospin factor $I^{\pm}$
since the $\pi^{-}/\pi^{+}$ ratio is almost a
constant in the absence of the Coulomb interactions.
The Coulomb impulse is
\begin{equation}
\delta\vec{p}=\pm(\vec{p}-\vec{p}_{0}(\vec{p}))
\end{equation}
for $\pi^{+}$ and $\pi^{-}$, respectively.
Thus, the impulse results in an increase (decrease) for
$\pi^{+}$ ($\pi^{-}$) in the $\sigma_{0}$ term of eq.\ (\ref{csigma})
which has approximately an exponential form $exp(-p^{2}_{0}/2m_{\pi}T)$,
where $T$ is the slope parameter of the pion spectra. The phase
space distortion $|\partial^{3}p_{0}/\partial^{3}p|$ generally results in a
decrease (increase) for $\pi^{+}$ ($\pi^{-}$), its exact form depends
on the dynamics of the nuclear charge distribution
and the trajectory of individual pions. An example of pion emission from the
origin of a sphere with the total charge Ze and a radius $r$ is
instructive, though not very realistic. The potential
energy of $\pm$ charged pions at the origin is $V_{0}
=\pm \frac{3}{2}Z\alpha/r$, energy conservation leads to
\begin{eqnarray}
\vec{p}_{0}(\vec{p})&=&\vec{p}(1\mp p^{2}_{c}/p^{2})^{1/2},
\end{eqnarray}
where $p_{c}=\sqrt{2m|V_{0}|}$.
One therefore obtains that
\begin{eqnarray}
|\frac{\partial^{3}p_{0}}{\partial^{3}p}|&=&(1\mp p^{2}_{c}/p^{2})^{1/2},
\end{eqnarray}
and
\begin{equation}
\sigma^{\pm}(p)=I^{\pm}\cdot\sigma_{0}(p)\cdot exp(\pm |V_{0}|/T)
(1\mp p^{2}_{c}/p^{2})^{1/2}.
\end{equation}
Keeping the first order in $Z\alpha$, the $\pi^{-}/\pi^{+}$ ratio
is approximately
\begin{equation}
R=\frac{\sigma(\pi^{-})}{\sigma(\pi^{+})}\approx R_{I}\cdot [1+2|V_{0}|
(\frac{m}{p^{2}}-\frac{1}{T})].
\end{equation}
The competition of the two factors $m/p^{2}$ and $1/T$
causes the $\pi^{-}/\pi^{+}$ ratio to change from $R\geq R_{I}$
to $R\leq R_{I}$ as the energy increases to above $E_{kin}=T/2$.
The variation of the $\pi^{-}/\pi^{+}$ ratio in the transport
model calculations is therefore understandable.

We now perform a comparison with the experimental data.
Before comparing the ratio we first compare in Fig.\ 2 and
Fig.\ 3 the $\pi^{+}$ and $\pi^{-}$
spectrum, respectively. The data were taken for the Au+Au reaction
at $\theta_{lab}=44^{0}$ by the KaoS collaboration at SIS/GSI\cite{muntzth}.
It is seen that the model can very well reproduce the available
experimental spectra. Fig.\ 4 shows a comparison of the $\pi^{-}/\pi^{+}$
ratio. The model calculation is in good agreement with the available
data at low energies, more complete data will be available
from both the KaoS and FOPI collaborations at SIS/GSI.
It is interesting to mention that a model calculation for the
reaction of Ni+Ni at $E/A=1.8$ GeV, which is almost
isosymmetric, shows that the $\pi^{-}/\pi^{+}$ ratio is about constantly 1
within statistical error bars. Comparing to the Au+Au reaction, it
further indicates the strong Coulomb effect on the $\pi^{-}/\pi^{+}$ ratio
in heavy systems.

In summary, using a hadronic transport model
we have shown that the recently observed large $\pi^{-}/\pi^{+}$
ratio at low energies in relativistic heavy ion collisions
is a manifestation of the strong Coulomb field of the reaction system.

The author is very grateful to C.M. Ko and G.Q. Li
for helpful discussions and their critical reading of the manuscript.
The author would also like to thank H. Oeschler, Ch. M\"untz, A. Wagner and
C. Sturm for interesting discussions.
The research was supported in part by NSF Grant No. PHY-9212209 and the Welch
Foundation under Grant No. A-1110.

\section*{Figure Captions}
\begin{description}

\item{\bf Fig.\ 1 }\ \ \ The calculated $\pi^{-}/\pi^{+}$ ratio
with and without Coulomb interactions for the reaction of Au+Au
at $E_{beam}/A=1.0$ GeV.

\item{\bf Fig.\ 2}\ \ \ Comparison to the $\pi^{+}$
spectrum measured at $\theta_{lab}=44^{0}\pm 4^{0}$ in the reaction
of Au+Au at $E_{beam}/A=1.0$ GeV

\item{\bf Fig.\ 3}\ \ \ Same as in Fig.\ 2 but for the preliminary
$\pi^{-}$ spectrum.

\item{\bf Fig.\ 4}\ \ \ The $\pi^{-}/\pi^{+}$ ratio for the
reaction of Au+Au at $E_{beam}/A=1.0$ GeV. The preliminary data was
measured at $\theta_{lab}=44^{0}\pm 4^{0}$ and the calculation is done at
$\theta_{lab}=45^{0}\pm 15^{0}$.
\end{description}

\end{document}